\documentclass[twocolumn,aps,showpacs,preprintnumbers,amsmath,amssymb]{revtex4}
\usepackage{array,graphicx,color,fancyhdr}
\usepackage{bm}% bold math

\begin{document}
\pagestyle{fancy}
\lhead{}
\lfoot{}
%\rfoot{}
%\chead{LA-UR-08-2796}
%\cfoot{LA-UR-08-2796}
\rhead{LA-UR-08-2796}
%\draft
%\preprint{W. T. Buttler and S. K. Lamoreaux, LA-UR-08-2796}
\title{\color{blue} Quantum cryptography with polarization, phase and time encoding}

\author{W. T. Buttler$^1$ and S. K. Lamoreaux$^2$}

\affiliation{$^1$Los Alamos National Laboratory \\
Physics Division (P-23), MS H803 \\
Los Alamos, NM 87545 \\ \\
$^2$Yale University \\
Physics - SPL, PO Box 208120 \\
New Haven, CT 06520-8120}

\begin{abstract}

We develop and present a quantum cryptography concept in which phase determinations are made from the time that a photon is detected, as
opposed to where the photon is detected, and hence is a non-interferometric
process. The phase-encoded quantum information is contained in temporal and polarization superpositions of single photon states, forming a complex qudit of {\it Hilbert} dimension $D \ge 4$. Based on this, we have developed a new quantum key distribution protocol that allows the generation of secret key in the presence of higher noise than is possible with other protocols.

\end{abstract}
\date{\today}
\pacs{03.67.Dd, 03.67.Hk, 42.50.Dv}
\maketitle

Bennett and Brassard proved the merit of Weisner's quantum cryptography (QC) foundations \cite{weisner} in 1984 with their presentation of the first mature quantum key distribution (QKD) protocol known as BB84 \cite{BB84}, which they experimentally demonstrated in 1989 \cite{bb84-exp}. Their protocol relies on transmission of a single photon onto a quantum communication channel by one party, Alice, to her partner with whom she wishes to share a secret, Bob.

The original scheme encodes information on a photon by preparing it in one of four definite polarizations: a horizontal or vertical polarized photon $ | h \rangle$ or $| v \rangle$ ($h$ or $v$), or a left- or right-diagonal polarized photon $| \bar{d} \rangle = (h - v)/\sqrt{2}$ or $| d \rangle = (h + v)/\sqrt{2}$ ($\bar{d}$ or $d$) (the early BB84 polarization protocol was investigated in fiber \cite{muller,breguet}). As the photons arrive to Bob, he randomly measures their polarizations in one of the two preparation bases, $h$-$v$ or $d$-$\bar{d}$. Half the time Bob measures the arriving states in the wrong basis giving random, meaningless results, but the other half of Bob's measurements are in the transmitted basis, determining the polarizations (bit values) of the arriving photons. Once the information exchange is complete Bob publicly reveals the times of his observations and his measurement basis. Alice then publicly notifies Bob which of his measurements were in the wrong basis and he discards that half of his bits to complete the sifting process. Bob and Alice then remove errors through a reconciliation protocol \cite{expcrypto,cascade,winnow,lodewyck}, and the exchange of the secret is completed through application of privacy amplification procedures \cite{expcrypto,privamp,winnow}. At a minimum, QC requires a QKD protocol, with sifting, reconciliation and privacy amplification protocols, but more procedures, such as signature authentication, may be required to ensure provable security.

Bennett extended QC to fiber-based phase-encoding schemes in 1992 \cite{b92} when he showed that QKD could be effected through use of a pair of Franson's unbalanced Mach-Zehneder interferometers (MZIs) \cite{franson91}. Fiber based BB84 phase-encoding QKD protocols typically encode photons with one of four phases, $\phi_i = 0$, $\pi$, $\pi/2$, or $3\pi/2$, and bit values are determined as 0 or 1 when Bob randomly adds a phase of 0 or $\pi/2$ to arriving photons, e.g., $\Delta \phi = 0 \equiv 0$, and $\Delta \phi = \pi \equiv 1$, (see \cite{townsend2}). The basis choice remains random yielding a theoretical protocol efficiency of $\eta_p = 1/2$ in a polarization switched system \cite{townsend1}.

In this manuscript we develop a QKD concept in which Bob transmits a temporal superposition of BB84 polarization states to Alice through an unbalanced MZI. For her part, Alice encodes phase information on the the arriving states with a phase shift of $\phi_i = 0$ or $\pi$ and then returns them to Bob. The polarization, phase and time encoding causes a complex qudit of high {\it Hilbert} dimension, allowing Bob to determine some phase-differences from the timing of photon detections, without entanglement or interferometric phase determination, to create a QC protocol of high efficiency and improved security.

The complex qudit presents with a high disturbance to eavesdropping, and maximizes the number of phase differences Bob can determine, which reduces the information an eavesdropper, Eve, can acquire. The reduction in information leakage implies that Bob and Alice can tolerate a higher error rate than presently known QKD protocols and still share a secret \cite{bruss,cerf}. The protocol is enabled by the encoding of four random phases of 0 or $\pi$ on the temporal and polarization superpositions, of the BB84 polarization states, through a phase-gate switch that breaks the {\it Faraday} mirror symmetry used in plug-and-play QKD \cite{gisin1,gisin2,bethune}.
\begin{figure}[b] %fig 1
\includegraphics[width=3.4 in,height=!,angle=0]{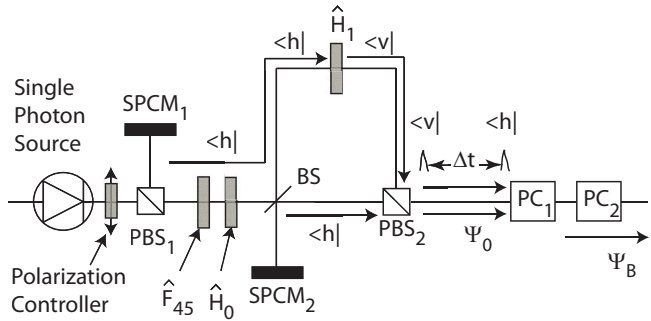}
\caption{Bob's optical system includes a single photon source, two single photon counting modules (SPCMs), two polarizing beamsplitters (PBSs), a $45^\circ$ Faraday rotator ($\widehat{F}_{45}$), two fixed-angle half-wave retarders ($\widehat{H}_0$ and $\widehat{H}_1$), a 50-50 beamsplitter (BS), and two Pockels cells (PC$_1$ and PC$_2$).} \label{transmitter}
\end{figure}

Figure \ref{transmitter} presents Bob's optics. His photon source is typically an attenuated laser source, but for the sake of argument we consider ideal single photon QKD. In this approach, a photon emitted from the source first encounters the polarization controller adjusted to transmit $h$. Next, the first polarizing beamsplitter (PBS$_1$) transmits $h$ toward the $45^\circ$ Faraday rotator, $\widehat{F}_{45}$, whose operation rotates $h$ to $d$. Following $\widehat{F}_{45}$ is $\widehat{H}_0$ that, when the photon is traveling this direction, causes a $45^\circ$ half-wave retardation that rotates $d$ back to $h$: $\widehat{H}_0 | d \rangle = h = \widehat{H}_0 \widehat{F}_{45} | h \rangle = h$, but when the photon returns, $\widehat{F}_{45} \widehat{H}_0 | h \rangle = v$, causing photons returning on this path to reflect to SPCM$_1$. This $h$ state is then superposed at the 50-50 beamsplitter (BS), with one superposition traveling the longer upper arm, and the other the shorter lower arm of the unbalanced MZI. The upper arm includes $\widehat{H}_1$ that causes a $90^\circ$ half-wave retardation that rotates $h$ to $v$: $\widehat{H}_1 | h \rangle = v$, and $\widehat{H}_1 | v \rangle = -h$ on the returning photons. The unbalanced MZI is completed with PBS$_2$ that transmits $h$ and reflects $v$ such that exiting the unbalanced MZI onto the quantum channel from Bob to Alice is $\Psi_0$, a dim-pulse superposition of $v$ delayed by $\Delta t$ following $h$,
\begin{eqnarray}
   \Psi_0 & = & | h \rangle \frac{g(\tau)}{\sqrt{2}} + i | v \rangle \frac{g(\tau + \Delta t)}{\sqrt{2}} \nonumber \\
      & \Leftrightarrow & \left [ \begin{array}{c} 1 \\ 0 \end{array} \right ] \frac{g(\tau)}{\sqrt{2}} + \left [ \begin{array}{c} 0 \\ i \end{array} \right ] \frac{g(\tau \! + \! \Delta t)}{\sqrt{2}}, \nonumber
\end{eqnarray}
where $i$ is the phase shift resulting from reflection from the BS, $\tau = t + z/c - t^\prime$ is the relative time from the beginning of Bob's transmission (the beginning of transmission is often heralded by Bob to Alice with a bright timing pulse at $t^\prime$), $\Delta t$ relates the length difference of the MZI short and long arms and is chosen long enough to allow operation of the Pockels cells (PCs) on both superpositions of $\Psi_0$ \cite{PC}, $g(\tau) = e^{i\omega \tau} \delta(\tau)$, $\delta(\tau)$ is a narrow envelope function with temporal width much greater that $1/\omega$ but less than $\Delta t$, and $c$ is the photon group velocity.

Assume the PCs transform the states as follows, where we choose $\alpha_i = 0, \pm \pi/4$, or $\pi/2$ to leave $h$ or $v$ unchanged, or to transform $h$ to $d$, $\bar{d}$ or $v$, or to transform $v$ to $d$, $\bar{d}$ or $h$. Consider the general case in which Bob applies rotation $\alpha_1$ to the early $h$, and rotation $\alpha_2$ to the late $v$ so that after the PCs Bob has prepared superposition $\Psi_B$ and sent it on to Alice:
\begin{equation}
   \Psi_B = \left [ \begin{array}{c} \cos \alpha_1 \\ \sin \alpha_1 \end{array} \right ] \frac{g(\tau)}{\sqrt{2}} + i \! \left [ \begin{array}{c} -\sin \alpha_2 \\ \cos \alpha_2 \end{array} \right ] \frac{g(\tau \! + \! \Delta t)}{\sqrt{2}}. \nonumber
\end{equation}
\begin{figure}[b] %fig 2
\vspace{-0.7 cm}
\includegraphics[width=3.25 in,height=!,angle=0]{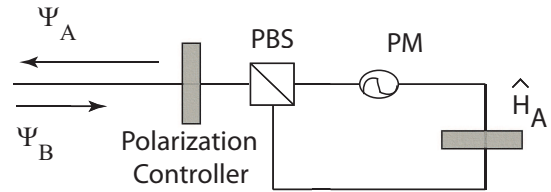}
\vspace{-1.0 cm} \caption{Alice's optical system is a phase-gate switch that includes an actively aligned polarization controller and a PBS followed by a loop with one phase modulator (PM), and a $90^\circ$ half-wave retarder, $\widehat{H}_A$ \cite{timing}. This phase gated switch does not rotate the polarizations $90^\circ$ as a {\it Faraday} mirror in plug-and-play \cite{gisin1,gisin2,bethune}.} \label{Alice}
\end{figure}

Figure \ref{Alice} shows Alice's optics \cite{timing}. As $\Psi_B$ arrives, her polarization controller corrects for the polarization rotations caused by the quantum channel, usually single-mode fiber, and her PBS then causes a spatial superposition of the arriving polarizations to form. For example, $h$ positions of $\Psi_B$ pass directly to phase modulator PM, and the $v$ positions reflect downward. The $h$ position of the first (earliest) temporal superposition encounters the PM first, and a phase $\phi_1 = 0$ or $\pi$ is added before it passes to $\widehat{H}_A$, which rotates it to $v$; the downward reflected $v$ position first passes through $\widehat{H}_A$ and is rotated to $-h$, and it is then given phase $\phi_2 = 0$ or $\pi$ as it passes through the PM; next, these two phase-modulated and polarization-switched superpositions recombine at the PBS at the earliest time position. A similar process occurs to the $i h$ and $i v$ positions of the later temporal position. That is, the $i h$ polarization passes directly through the PBS to the PM that adds phase $\phi_3 = 0$ or $\pi$ to the position, and it is then rotated to $i v$ by $\widehat{H}_A$; the $i v$ position is reflected downward by the PBS, rotated to $-i h$ by $\widehat{H}_A$, and given $\phi_4 = 0$ or $\pi$ by the PM; the phase-modulated and polarization-switched superpositions then recombine at the PBS at the latest time position of the temporal superposition to form $\Psi_A$,
\begin{equation}
   \Psi_A = \left [ \! \begin{array}{c} -e^{i \phi_2} \sin \alpha_1 \\ e^{i \phi_1} \cos \alpha_1 \end{array} \! \right ] \frac{g(\tau)}{\sqrt{2}} - i \! \left [ \! \begin{array}{c} e^{i \phi_4} \cos \alpha_2 \\ e^{i \phi_3} \sin \alpha_2 \end{array} \! \right ] \frac{g(\tau \! + \! \Delta t)}{\sqrt{2}}, \nonumber
\end{equation}
which Alice directs toward the polarization controller. The controller then inverts its earlier rotations before directing the temporal, polarization, and phase modulated qudit onto the quantum channel to Bob. It is a subtle point, but, if Bob sent to Alice a state that includes a $d$ or $\bar{d}$ within the superposition, after passing through the phase-gate switch that position appears to have been reflected back to him from a mirror: $d \mapsto \bar{d}$ and $\bar{d} \mapsto d$. However, if his superpositions included $v$ or $h$, those positions will appear to have been switched, in his coordinate system: $h \mapsto v$ and $v \mapsto h$. In contrast, if Alice's optics behaved as a {\it Faraday} mirror, e.g., the operations on the states effect a $90^\circ$ rotation in all cases: $h \mapsto \pm v$, $d \mapsto \pm \bar{d}$, $v \mapsto \mp h$, and $\bar{d} \mapsto \mp d$, from Bob's perspective.

As $\Psi_A$ arrives to Bob, he appropriately modulates his two PCs to select $\alpha_3$ and $\alpha_4$ to determine most of the phases Alice has added to the $h$ and $v$ polarizations to form the generalized state, $\Psi^\prime_B$ after his PCs and prior to PBS$_2$:
\begin{eqnarray}
   \Psi^\prime_{B} \! & \! = \! & \! - \! \! \left [ \begin{array}{c} e^{i \phi_2} \sin \alpha_1 \cos \alpha_3 + e^{i \phi_1} \cos \alpha_1 \sin \alpha_3 \\ e^{i \phi_2} \sin \alpha_1 \sin \alpha_3 - e^{i \phi_1} \cos \alpha_1 \cos \alpha_3 \end{array} \right ] \frac{g(\tau)}{\sqrt{2}} \nonumber \\ \nonumber \! \! - \! \! \! &i& \! \! \! \! \left [ \begin{array}{c} e^{i \phi_4} \cos \alpha_2 \cos \alpha_4 - e^{i \phi_3} \sin \alpha_2 \sin \alpha_4 \\ e^{i \phi_4} \cos \alpha_2 \sin \alpha_4  + e^{i \phi_3} \sin \alpha_2 \cos \alpha_4 \end{array} \right ] \frac{g(\tau \! + \! \Delta t)}{\sqrt{2}}. \nonumber
\end{eqnarray}
Then, as $\Psi^\prime_B$ enters the MZI, PBS$_2$ superposes the early and late time positions, based on polarizations, to form yet another temporal superposition. For example, at the earliest time position the $h$ position passes directly through to time $\tau + 0$, and the $v$ position travels the upper arm to time $\tau + \Delta t$; similarly, at the latest time position the $h$ positions pass directly through to time $\tau + \Delta t$, and the $v$ position travels the upper arm to time $\tau + 2 \Delta t$. The photon will be detected on coherence at one of these three times, on one of the two SPCMs, based on Bob's decoding operations.

To complete a QKD protocol, we consider three general classes of PC rotations. The first class is when Bob chooses $(\alpha_1,\alpha_2) = (0,0)$, $(0,\pi/2)$, $(\pi/2,0)$, or $(\pi/2,\pi/2)$. In this situation, Bob can overlap and interfere his states with efficiency $\eta = 1$ at time $\tau + \Delta t$ by choosing $(\alpha_3,\alpha_4) = (\alpha_1,\alpha_2)$ to determine $\Delta \phi_{41}$, $\Delta \phi_{31}$, $\Delta \phi_{42}$, and $\Delta \phi_{32}$, respectively, on one of his two SPCMs. After Bob makes these phase difference determinations he notifies Alice of the relative time $\tau$ (the absolute time that includes $\Delta t$ is not needed) of the observation and his measurement basis ($\Delta \phi_{41}$, $\Delta \phi_{31}$, $\Delta \phi_{42}$, or $\Delta \phi_{32}$).

The second class is when Bob chooses $(\alpha_1,\alpha_2) = (0,\pi/4)$, $(\pi/2,\pi/4)$, $(0,-\pi/4)$, $(\pi/2,-\pi/4)$, $(-\pi/4,0)$, $(-\pi/4,\pi/2)$, $(\pi/4,0)$, or $(\pi/4,\pi/2)$. In these situations, Bob chooses $\alpha_3 = \pi/2$ if $\alpha_1 = 0$, and $\alpha_3 = 0$ if $\alpha_1 = \pi/2$, to cause an early-time coherence of $h$ or $v$ of his superposition to randomly trigger either SPCM$_1$ or SPCM$_2$ at the earliest possible detection time $\tau + 0$, without interference and no phase difference determination. Similarly, Bob chooses $\alpha_4 = \pi/2$ if $\alpha_2 = 0$, and $\alpha_4 = 0$ if $\alpha_2 = \pi/2$, to cause a late-time coherence of $i h$ or $i v$ of his superposition to randomly trigger either SPCM$_1$ or SPCM$_2$ at the latest possible detection time $\tau + 2 \Delta t$, without interference or any phase determination. In the first situation (early-time $h$ or $v$ transmission, i.e., $\alpha_1 = 0$ or $\pi/2$), Bob chooses $\alpha_4 = \pi/4$, and in the latter situation (late-time $i h$ or $i v$ transmission) Bob chooses $\alpha_3 = \pi/4$. Bob ignores the 1/2 of these detections that are random as they convey no phase information. For the remaining cases Bob finds with $\eta = 1/2$ that the polarization of his photon, and thus the time of an observation, depends on $\Delta \phi_{21}$ or $\Delta \phi_{43}$, respectively, and he notifies Alice of the relative time $\tau$ of his detection and the measurement basis. For example, consider that Bob chose $(\alpha_1,\alpha_2,\alpha_3,\alpha_4) = (0,\pi/4,\pi/2,\pi/4)$. This gives
\begin{equation}
   \Psi^\prime_B = - \! \! \left [ \begin{array}{c} e^{i \phi_1} \\ 0 \end{array} \right ] \frac{g(\tau)}{\sqrt{2}} -i \! \left [ \begin{array}{c} e^{i \phi_4} - e^{i \phi_3} \\ e^{i \phi_4} + e^{i \phi_3} \end{array} \right ] \frac{g(\tau + \Delta t)}{2 \sqrt{2}} \nonumber
\label{timing} \end{equation}
prior to PBS$_2$ demonstrating that the earliest detections after $\Psi^\prime_B$ passes through the MZI are indeterminate, but any detection on any SPCM at time $\tau + \Delta t$ determines that $\Delta \phi_{43} = \pm \pi$, and that a detection on any SPCM at time $\tau + 2 \Delta t$ determines that $\Delta \phi_{43} = 0$. Thus, Bob determines Alice's phase differences by {\it when} he detected a photon, the time of the event, rather than {\it where} he detected a photon, or interferometric determination \cite{options}.

The third class to consider is when Bob chooses $(\alpha_1,\alpha_2) = (\pi/4,\pi/4)$, $(\pi/4,-\pi/4)$, $(-\pi/4,\pi/4)$, or $(-\pi/4,-\pi/4)$. This class is similar to the second in that he determines phase differences based on the time of detections rather than by interferometric means. He chooses his rotations to determine $\Delta \phi_{21}$ at time $\tau + 0$, and $\Delta \phi_{43}$ at time $\tau + 2 \Delta t$ with efficiency $\eta = 1/2$. For example, suppose that Bob chose $(\alpha_1,\alpha_2,\alpha_3,\alpha_4) = (\pi/4,-\pi/4,\pi/4,\pi/4)$. This gives
\begin{equation}
   \Psi^\prime_B = - \! \! \left [ \begin{array}{c} e^{i \phi_2} + e^{i \phi_1} \\ e^{i \phi_2} - e^{i \phi_1} \end{array} \right ] \frac{g(\tau)}{2 \sqrt{2}} - i \! \left [ \begin{array}{c} e^{i \phi_4} + e^{i \phi_3} \\ e^{i \phi_4} - e^{i \phi_3} \end{array} \right ] \frac{g(\tau + \Delta t)}{2 \sqrt{2}} \nonumber
\label{timing2} \end{equation}
prior to PBS$_2$. Detections at time $\tau + 0$ determine $\Delta \phi_{21} = 0$, and detections at time $\tau + 2 \Delta t$ determine $\Delta \phi_{43} = \pm \pi$, while detections at time $\tau + \Delta t$ are not determined, or are effectively random \cite{options}. As usual, Bob notifies Alice of the relative time $\tau$ and phase basis of his determination.

With these physics Bob and Alice build a QKD protocol that allows determination of $\Delta \phi_{21}$, $\Delta \phi_{31}$, $\Delta \phi_{41}$, $\Delta \phi_{32}$, $\Delta \phi_{42}$, or $\Delta \phi_{43}$. The protocol is most easily balanced by Bob sending one of the $(h,i h)$, $(h,i v)$, $(v,i h)$, or $(v,i v)$ states as often as he sends any of the remaining twelve polarization combinations giving $\eta_p = 0.75$, but Alice and Bob could also balance the protocol by extending the class-two and class-three degrees of freedom, with a lower $\eta_p$, by choosing $\alpha_3$ and $\alpha_4$ to interferometrically determine $\Delta \phi_{31}$, $\Delta \phi_{32}$, $\Delta \phi_{41}$, or $\Delta \phi_{42}$ with $\eta = 1/2$ at time $\tau + \Delta t$.

A complete error analysis of this protocol will evaluate the qudit security against individual \cite{fuchs}, coherent \cite{Lo2,shor,tittel}, and cloning \cite{buzek1,buzek2,werner} attacks. These attacks consider the {\it Hilbert} dimension $D$ and the number of conjugate bases. For our case, Bob has prepared and sent a qudit $\Psi_B$ with $D = 4$, which was shown in \cite{bruss,cerf} that it can be secure for disturbances of up to 25\% \cite{point2point}. As Bob's qudit arrives Alice then switches the polarizations and adds four phases to the qudit prior to its return to Bob, increasing the allowable disturbance further.

These strong attacks typically require Eve to store her probe until {\it after} Bob and Alice publicly announce their measurement basis, implying that if Alice and Bob simply encrypt their sifting communications, or even simply wait some time to sift their quantum communication, that they can guard against these attacks. Nevertheless, these strategies define the strictest bounds on the key rates of QKD protocols against generalized attacks. Upon defeat, however, a reasonable attack by Eve is the intercept-resend strategy, and it should be evaluated as well.

Due to the number of degrees of freedom in the qudit, Alice and Bob have many choices that have all been shown to be more secure against eavesdropping \cite{bruss,cerf}. They could even effect free-space QKD \cite{bb84-exp,franson-fs,buttler,butt98-prl,butt00-prl} with Bob's qudit alone, with enhanced security and equivalent efficiency as BB84.

In summary, we have developed and analyzed three physics concepts that enhance the security of quantum cryptography. First, we developed (Bob's) optics that exploit polarization and time encoding to form a complex qudit with $D = 4$, which presents with a high tolerance to eavesdropping. Next, we fashion a phase-gate switch that breaks the plug-and-play {\it Faraday} mirror symmetry by switching incident polarizations in a manner that allows addition of a single relative phase to arriving polarization superpositions, increasing the security of the qudit further against eavesdropping. Finally, we presented the complex qudit realization of differential phase determination without interferometric means. These elements can be used to establish a QKD protocol that exploits time, phase, and polarization encoding, providing significantly improved immunity to errors, and some of the elements may have broader uses in the general quantum information field. In addition, the security  {\it check} is better than any other protocol described to date. In our presentation, we have assumed a single photon source. In the case of a weak coherent source, the security will of course be reduced as in all QKD systems.

%\clearpage

\clearpage


\begin{thebibliography}{100}
\bibitem{PC} On transmission, Bob uses only one PC to create $\Psi_B$. During his analysis Bob uses both PCs to decode the states. The PCs have the fast and slow axes conjugately aligned.
\bibitem{timing} Not shown are synchronizing optics Alice uses to common time her phase modulations with Bob's transmissions. Timing could be effected with a single bright pulse to herald the beginning of Bob's transmission, on a separate fiber or the same fiber with wavelength-division multiplexing, and quality local oscillators at both Alice's and Bob's locations.
\bibitem{options} Alice and Bob can vary their $(\alpha_3,\alpha_4)$ choice so that detections at $\tau + 0$ determine $\Delta \phi_{21} = \pm \pi$, and detections at $\tau + 2 \Delta t$ determine $\Delta \phi_{43} = 0$.
\bibitem{point2point} Bob and Alice could use the polarization and time encoded qudit to effect direct point-to-point QKD.
\end{thebibliography}
\end{document}